\documentclass[aps,pra,reprint,twocolumn,showpacs,floatfix,superscriptaddress]{revtex4-2}

\usepackage{amssymb,amsmath,amstext}                
\usepackage{graphicx}                                               
\usepackage{epstopdf}                                               
\usepackage{color}                                                     
\usepackage{bm}                                                        
\usepackage{appendix}                                              
\usepackage[utf8]{inputenc}
\usepackage{latexsym}
\usepackage{xcolor}
\usepackage{braket}
\usepackage{ulem}
\usepackage{siunitx}
\usepackage{umoline}
\usepackage{epsfig}
\usepackage{graphics}
\usepackage{psfrag}
\usepackage{mathtools}
\usepackage[english]{babel}
\usepackage[T1]{fontenc}
\usepackage{lmodern}
\usepackage{booktabs}
\definecolor{Maroon} {RGB}{219, 48, 122}

\newcommand{\ad}[1]{{\color{Maroon}{{#1}}}}

\usepackage{rotating}
\usepackage{multirow}

\definecolor{lblue} {RGB}{51,71,158}
\usepackage[colorlinks=true,citecolor=blue,linkcolor=blue,urlcolor=lblue]{hyperref}

\usepackage{tikz}
\definecolor{lime}{HTML}{A6CE39}
\DeclareRobustCommand{\orcidicon}{%
    \begin{tikzpicture}
    \draw[lime, fill=lime] (0,0) 
    circle [radius=0.13] 
    node[white] {{\fontfamily{qag}\selectfont\tiny ID}};
    \draw[white, fill=white] (-0.0625,0.095) 
    circle [radius=0.007];
    \end{tikzpicture}
    \hspace{-2mm}}
    
\newcommand{\orcidJK}{\href{https://orcid.org/0000-0003-0998-9460}{\orcidicon}}
\newcommand{\orcidMP}{\href{https://orcid.org/0009-0009-1858-882X}{\orcidicon}}
\newcommand{\orcidAA}{\href{https://orcid.org/0000-0002-4776-1523}{\orcidicon}}

\begin{document}
\title{Interplay of localization and topology \\ in disordered  dimerized array of Rydberg atoms }

\author{Maksym Prodius\orcidMP}
    \email{maksym.prodius@uj.edu.pl}
    \affiliation{Szko\l{}a Doktorska Nauk \'Scis\l{}ych i Przyrodniczych, Uniwersytet Jagiello\'nski, ulica Stanis\l{}awa \L{}ojasiewicza 11, PL-30-348 Krak\'ow, Poland}
 \affiliation{Instytut Fizyki Teoretycznej, Wydzia\l{} Fizyki, Astronomii i Informatyki Stosowanej, Uniwersytet Jagiello\'nski, \L{}ojasiewicza 11, PL-30-348 Krak\'ow, Poland} 

\author{Adith Sai Aramthottil\orcidAA}
 \affiliation{Instytut Fizyki Teoretycznej, Wydzia\l{} Fizyki, Astronomii i Informatyki Stosowanej, Uniwersytet Jagiello\'nski, \L{}ojasiewicza 11, PL-30-348 Krak\'ow, Poland}

\author{Jakub Zakrzewski\orcidJK}
\email{jakub.zakrzewski@uj.edu.pl}
\affiliation{Instytut Fizyki Teoretycznej, Wydzia\l{} Fizyki, Astronomii i Informatyki Stosowanej, Uniwersytet Jagiello\'nski, \L{}ojasiewicza 11, PL-30-348 Krak\'ow, Poland} 
\affiliation{Mark Kac Complex Systems Research Center, Jagiellonian University in Krak\'ow, PL-30-348 Krak\'ow, Poland} 

\date{\today} 

\begin{abstract}

Rydberg tweezer arrays provide a platform for realizing spin-1/2 Hamiltonians with long-range tunneling that decays as a power law with distance. We numerically investigate the effects of positional disorder and dimerization on the properties of excited states in such a one-dimensional system. Our model allows for continuous tuning of both the dimerization pattern and the disorder strength. Within the parameter space constrained by our geometry, we show that both mechanisms lead to a localized phase that does not resemble standard MBL. This phase can be understood as an ensemble of distinct Hilbert-space–fragmented realizations induced by small inter-spin separations. As dimerization is commonly associated with  Symmetry Protected Topological (SPT) physics, we also examine the SPT states across the entire energy spectrum. Despite the presence of a partial spin-glass order, we argue that the system hosts an extensive fraction of SPT states.

\end{abstract}


\maketitle

\section{Introduction} \label{sec:intro}

Interacting quantum many-body systems typically follow the Eigenstate Thermalization Hypothesis (ETH) \cite{Deutsch91,Srednicki94, Srednicki99}.
Intensively studied over the last 20 years many-body localization (MBL)
is a robust example of ergodicity breaking phenomenon (see early works
\cite{Fleishman80, Gornyi05, Basko06, Oganesyan07} as well as several reviews on the topic
\cite{Nandkishore15, Alet18, Abanin19, Sierant25}). Typically occurring due to the presence of a strong disorder, MBL phase is believed to be integrable, characterized by a set of local integrals of motion (LIOMs) \ad{\cite{Serbyn13a,Serbyn13b,Chandran15,Imbrie17}} and area-law entangled eigenstates. The lack of thermalization for such systems was demonstrated experimentally \cite{Schreiber15,Smith16} showing that the memory of the initial state is indeed preserved for achievable times. 

Later other manifestations of nonergodicity, not necessarily related to the standard MBL, were discovered such as, e.g., many-body quantum scars \cite{Shiraishi17, Bernien17,Turner18,Serbyn21,Moudgalya22rev} or dynamics in systems with Hilbert space fragmentation, typically associated with the (approximate) preservation of some, not necessarily local constants of the motion \cite{Khemani20,Rakovszky20,Li21}. In particular, tilted one-dimensional models reveal the fragmentation due to emerging conservation of the global dipole moment \cite{vanNieuwenburg19,Schulz19, Taylor20, Khemani20,Rakovszky20, Yao20b, Yao21} even in the absence of disorder. For the latter systems, the localization mechanism differs from that in the standard MBL, in particular, as discussed in~\cite{Rakovszky20}, instead of LIOMs one may consider their possible non-local generalizations (for a recent discussion see~\cite{Lisiecki25}). Hilbert space {fragmentation} based localization has also been observed 
experimentally \cite{Guo20, Kohlert23,Scherg21}.

As {another} fascinating idea,
it has been suggested \cite{Huse13}  that the quantum order of ground state, such as Spontaneous Symmetry Breaking (SSB), Topological Order or Symmetry Protected Topological (SPT) order can be promoted to the excited spectrum in an area-law entangled phase, such as MBL. Shortly afterwards, few such systems were indeed found \cite{Chandran14, Bahri15,  Vasseur16, Parameswaran18, Kuno19,Leseleuc2019,Mondal22,
Laflorencie22, Brighi25}. Interestingly, however, the existence of strong edge modes has also been postulated for some Hilbert space fragmented models \cite{Rakovszky20}. 

 In this work, we study the properties of the excited states in a long-range, bond-disordered XY model  - an experimentally accessible setup based on Rydberg atoms in tweezer arrays \cite{Browaeys20}. The dipolar coupling between atoms depends on the distance between them (as well as their orientation), thus one can realize such chains by arranging the tweezers with positional randomness, or incorporate such disorder along other positional configurations. In particular, we investigate how the localization properties of the model are affected by the combined presence of disorder and dimerization of spin positions. We show that these two mechanisms give rise to an ergodicity-breaking regime characterized by a partial spin-glass order \cite{Parisi80,Parisi83}, but which does not resemble standard many-body localization (MBL). Moreover, this system can be viewed as an ensemble of disordered Hamiltonians that exhibit distinctive patterns of Hilbert-space fragmentation determined by the particular disorder realizations. As shown by some of us recently \cite{Aramthottil24} (see also \cite{Zhao25}) such
bond-disordered systems can be treated within the framework of the real-space renormalization group for excited states (RSRG-X) \cite{Pekker14}. RSRG-X captures the observed multimodal entanglement entropy distributions and sub-Poissonian mean gap ratio observed in such random-bond models \cite{Aramthottil24,Braemer22}, the latter property has been observed in the Hilbert space fragmented setting \cite{Jeyaretnam25}.  The integrals of motion for the bond-disordered models are rather the non-local generalizations discussed in the Hilbert space fragmentation setting~\cite{Rakovszky20,Lisiecki25}, than standard LIOMs~\cite{Aramthottil24}.

As dimerization is one of the simplest routes into SPT physics \cite{Su79}, we examine the interplay between localization and topology in our setup. Despite indications of a partial  spin-glass order, we investigate whether  nontrivial SPT states can coexist with glassy states and, in doing so, we discuss the challenges associated with their numerical detection.

\begin{figure}[ht]
    \centering
    \includegraphics[width=.98\linewidth]{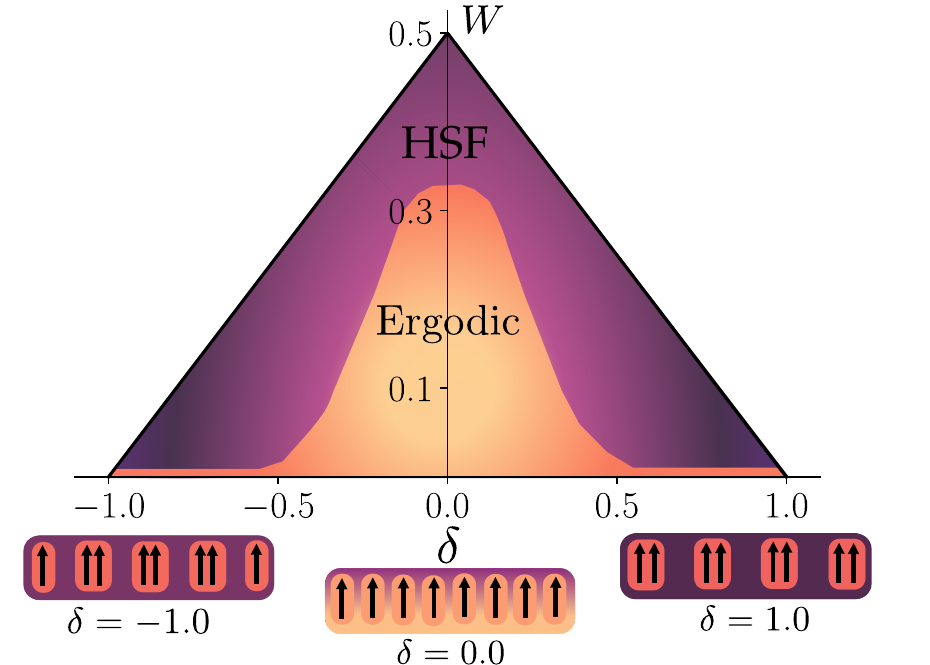} 
    \caption{The qualitative phase diagram for the model in $\delta,W$ space where $\delta$ is the dimerization parameter in the model defined by \eqref{eq:model} and \eqref{eq:pos} and $W$ is the positional disorder strength. The ergodic domain at low $W,|\delta|$ is surrounded by a localized region characterized by Hilbert space fragmentation (HSF). The bottom panels visualize spin positions for different dimerization, $\delta$.}
    \label{fig:schem}
\end{figure}

Our work is organized as follows. In the next Section, we introduce the considered model. The statistical properties of eigenvalues and eigenvectors are discussed in detail in Section~\ref{sec:stat}. Subsequently, we present the spin-glass order analysis in Section~\ref{sec:gl} and the renormalization-group understanding of the model in Section~\ref{RG}, followed by the analysis of its topological properties in Section~\ref{sec:SPT}. Possible experimental verifications in the time dynamics of well-chosen initial states are described in Section~\ref{sec:time}, and we conclude in Section~\ref{sec:conc}. The appendices present additional details.

\section{The model}
\label{sec:model}

For simplicity, we study a one-dimensional spin-1/2 lattice model inspired by experiments with arrays of Rydberg atoms kept in tweezers \cite{Browaeys20}, where, e.g., the first realization of a many-body SPT phase of interacting bosons in an artificial system was recently reported \cite{Leseleuc2019}. 
We concentrate on the XY model as realized for atoms due to direct dipole-dipole interactions between Rydberg states of different parity \cite{Browaeys20,Chen23,Emperauger25} and we consider a one-dimensional (1D) chain.
Using a Rydberg tweezer platform, one may realize different dimerization patterns combined with positional disorder, enabling access to different SPT phases and ergodic/ergodicity-breaking regimes.

The model discussed is defined by the Hamiltonian 
\begin{equation}
    \label{eq:model}
         H = J\sum_{i>j} \frac{1}{r_{ij}^3} \left( s_i^x s_j^x + s_i^y s_j^y\right),
\end{equation}
where $J=1$ sets the energy unit (we use later a notation $J_{ij}\equiv J/r^3_{ij}$) and {$s_i^k$, $k=x,y,z$} are spin-1/2 operators. Spins are placed at positions $r_i$ with  $r_{ij} \equiv  \left|r_i - r_j\right|$. We assume open boundary conditions (OBC).

Hamiltonian (\ref{eq:model}) obeys $U(1)$ symmetry, which corresponds physically to the conservation of a total magnetization along the $z$ direction. 
The second symmetry, crucial for the protection of SPT order, is represented by an antiunitary $S = \prod_{i} s_{i}^x \circ K$, where $K$ is a complex conjugation. This makes the symmetry group of our system $U(1) \times \mathbb{Z}_2^{T}$. Note that the coupling between spins 
is long-range (decaying as a power law) 
so we deal with the bosonic interacting SPT system \cite{Chen11, Chen11com}. The interacting character of the model may also be visualized in the fermionic language via the Jordan-Wigner transformation, where long-range coupling results in terms quartic in fermionic operators \cite{Leseleuc2019}.

We introduce disorder into our model by randomly repositioning spins. Backed by SSH logic~\cite{Su79}, 
we distribute the positions of spins as
\begin{equation}
    r_i= \left(i + \frac{\delta(-1)^{i+1}-1}{2} \right)a + i \hspace{0.3ex}D_{rb}  + W_i,
    \label{eq:pos}
\end{equation}
where parameter $\delta \in \left[ -1, 1 \right]$ quantifies the degree of dimerization. We have the chain with paired spins in the bulk and two almost decoupled spins at the edges for $\delta=-1$ and all spins paired for $\delta = 1$ - see Fig.~\ref{fig:schem}, 
{the corresponding regions are loosely called nontrivial and trivial, respectively}. Random $W_i$ values are drawn from the uniform distribution $W_i\in[-W,W]$ and $D_{rb}$ is the  Rydberg blockade diameter~\cite{Lukin01}. Therefore, $\tilde{a}=a+{D}_{rb}$ gives the average distance between two neighboring spins. From now on, we fix $a=1$ and ${D}_{rb}=0.2$.

Assuming the straight line chain geometry we obtain the constraint on $W$ and $\delta$: 
\begin{equation}
\label{eq:constrain}
         W \hspace{0.3ex} <  \hspace{0.3ex}  a \cdot \min\left(\frac{1-\delta}{2},\frac{\delta + 1} {2}\right),
\end{equation}
leading, for fixed $a$ and $D_{rb}$, to a triangular parameter space -- compare Fig.~\ref{fig:schem}. The system properties change considerably, depending on $W,\delta$ parameters as we discuss below.

\section{Statistical analysis of spectra}
 \label{sec:stat}

To characterize the properties of the system for different parameter values, we use 
the standard tools, such as the gap ratios and the half-chain entanglement entropies (EE) of eigenstates. The gap ratio is defined as $r_k = \frac{\min\lbrace \delta_k, \delta_{k+1} \rbrace}{\max\lbrace \delta_k, \delta_{k+1} \rbrace}$, where $\delta_k = E_{k+1} - E_k$ is the difference between two consecutive eigenvalues of (\ref{eq:model}). The mean gap ratio is found as the average value taken over at maximum $1000$ eigenvalues in the middle of the spectrum for a given disorder realization.
We work in the zero-magnetization and even-parity sector across the paper. 
We take between $100$ for $L = 20$ and $5000$ for $L=12$ different disorder realizations. The half-chain EE is defined as
$S_{ent}(L/2) = -\mathrm{Tr}(\rho \log_2(\rho))$ (where $\rho$ is the density matrix obtained by tracing out the eigenstate over half of the chain). Again, the mean is taken over eigenstates in the middle of the spectrum. In the shown plots, we present this mean EE, $S_{ent}$, normalized by the Page value, $S^P$,
i.e. a typical entanglement entropy of the random state \cite{Page93average}.

\begin{figure}[ht]
    \centering
    \includegraphics[width=1.0\linewidth]{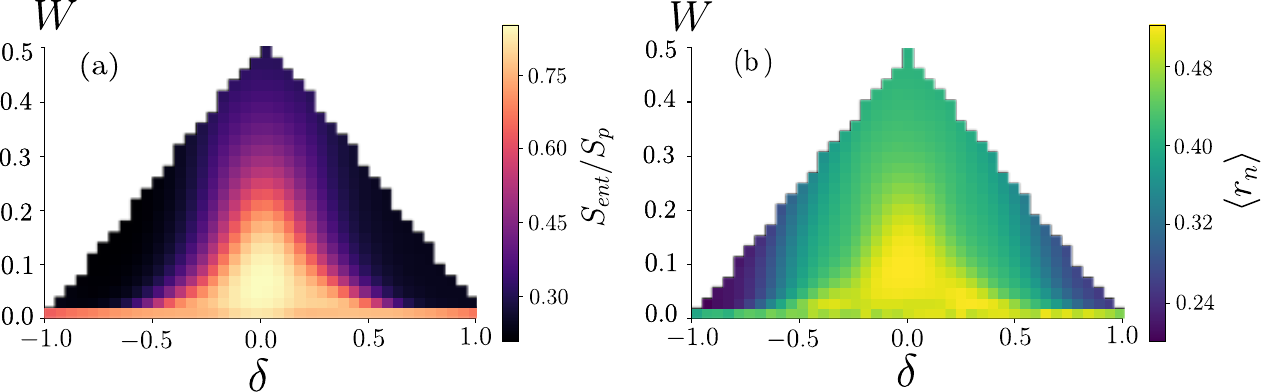}
\caption{Half-chain entanglement entropy $S_{ent}/S^P$ normalized by the Page value (a) and the mean gap ratio $\braket{r_n}$ (b) as a function of the disorder strength $W$ and the dimerization, $\delta$. The ergodic area in the middle, for $\delta>0$, is surrounded by localized regions characterized by low entanglement entropy as well as mean gap ratio being Poissonian or even strongly sub-Poissonian close to borders of the space of parameters.}
    \label{fig:gap_ratio_ent_triangles}
\end{figure}

To have a rough idea of the localization properties of the system, we make a scan of the whole triangular $W,\delta$ parameter space for a relatively small 
system size $L=14$ - c.f. Fig.~\Ref{fig:gap_ratio_ent_triangles}. 
Clearly, while for $W=0$ the whole line of the dimerization parameter, $\delta$, corresponds to the delocalized case; in the presence of disorder the situation is different. We observe two main regions: the delocalized, ergodic region in the center, for small $W$ and $\delta$ is surrounded by an apparently localized domain for sufficiently large $W$ or $|\delta|$. A careful observer will notice that the mean gap ratio is slightly above the Poissonian value $r_{\rm Poi}=0.389$ for large $W$, small $\delta$ region, while for small disorder and large magnitude of dimerization, $|\delta|$, it becomes significantly sub-Poissonian. We shall analyze this behavior in detail below.   

\begin{figure}[ht]
    \centering
    \includegraphics[width=1.0\linewidth]{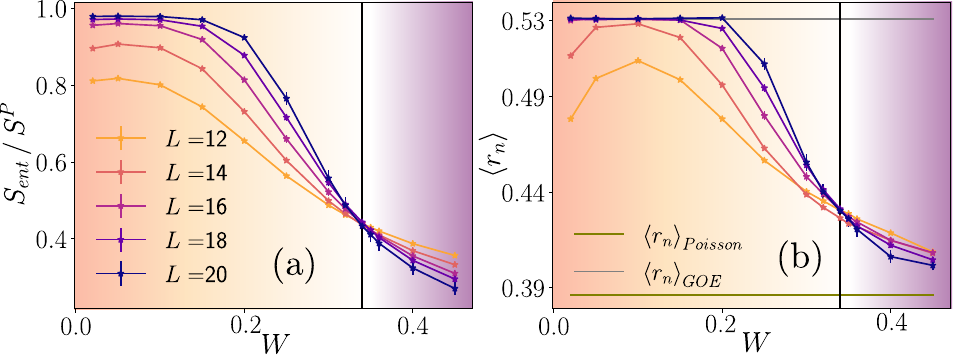}
     \centering
    \includegraphics[width=1.0\linewidth]{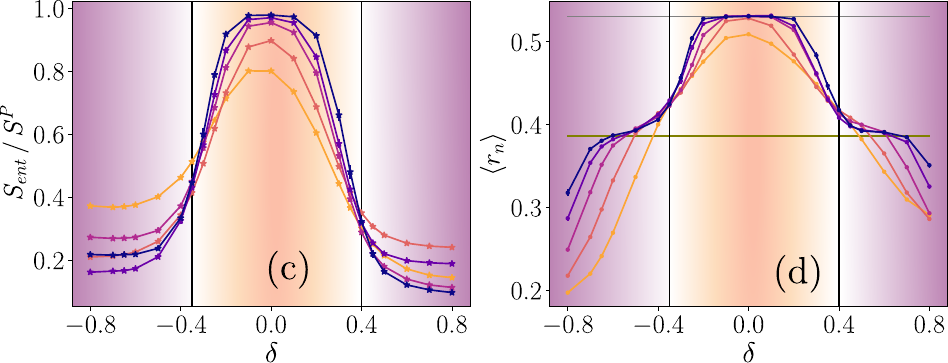}
\caption{Half-chain entanglement entropy $S_{ent}$  and the mean gap ratio $\braket{r_n}$  as a function of  the disorder strength $W$ for fixed $\delta = 0$ (a-b) and as a function of dimerization $\delta$ for  fixed $ W= 0.1$ (c-d).}
    \label{fig:entropy_gap_ratio}
\end{figure}

First, we benchmark the localization crossover depending on the disorder strength $W$ for a uniform pattern of spins ($\delta = 0$, along a vertical line in Fig.~\ref{fig:schem}).
The results are plotted in panels (a) and (b) in 
 Fig.~\ref{fig:entropy_gap_ratio} for different system sizes. To diagonalize the Hamiltonian we employ an exact diagonalization for $L \le 16$ and POLFED algorithm \cite{Sierant20polfed} for larger system sizes. One observes that both the mean EE and the mean gap ratio curves for different system sizes cross around $W \approx 0.33$, indicating a crossover between the delocalized and localized behavior. A careful observer will notice significant size effects. Only for $L\ge 14$ for small disorder values, the expected ergodic dynamics with mean gap ratio corresponding to the Gaussian orthogonal ensemble (GOE)  are reached. 
 On the localized side, even for the largest disorder, the Poissonian value {$r_{\mathrm{Poi}}$} for $\langle r_k \rangle$ is not reached, staying above 0.4. These conclusions are mirrored by the mean rescaled EE. The curves for different system sizes cross nicely around the same disorder value, $W \approx 0.33$ showing, in fact, a {volume-law} scaling on the delocalized side. In contrast, for larger disorder sub-volume behavior is reflected by the inverse (as compared to the delocalized side) ordering of curves corresponding to different system sizes. 
 
Subsequently, we fix the disorder strength at $W=0.1$, corresponding to a delocalized regime for uniformly spaced spins, and vary the dimerization parameter $\delta$ - c.f. Fig.~\ref{fig:entropy_gap_ratio}(c-d).  For small $\delta$, the system is delocalized with the mean gap ratio and the mean EE reaching values corresponding to GOE for systems of sufficient size. The mean EE and the mean gap ratio reveal two crossovers for $\delta$ sufficiently different from $0$ occurring for small disorder $W$ values. A careful observer will notice
that curves in Fig.~\ref{fig:entropy_gap_ratio}(c) in localized regimes are not ordered by the system size as seen in other panels, where, typically, a monotonic dependence of a given quantity on the system size is observed. 
This unusual behavior on the localized side in Fig.~\ref{fig:entropy_gap_ratio}(c) occurs because we cut the system for EE calculations in the middle of the chain, i.e., on the strong/weak bond, depending on $L$. 
The crossover between localized and delocalized phases occurs, for 
 $\delta \approx -0.35$ and   $\delta \approx 0.4$ as indicated by the crossing of curves. The phases for large $|\delta| $ are localized as indicated by the mean gap ratio and sub-volume mean EE. Note, however, that the mean gap ratio on the localized side, instead of being close to the Poissonian value,
characteristic to a ``standard MBL'' case, becomes strongly sub-Poissonian.
This resembles the observations made in  \cite{Decker20}).

\begin{figure}[ht]
    \centering
    \includegraphics[width=1.0\linewidth]{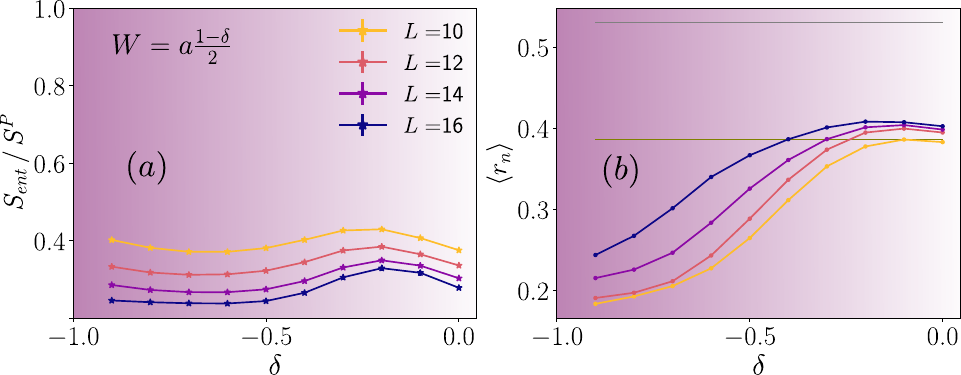}
     \centering
    \includegraphics[width=1.0\linewidth]{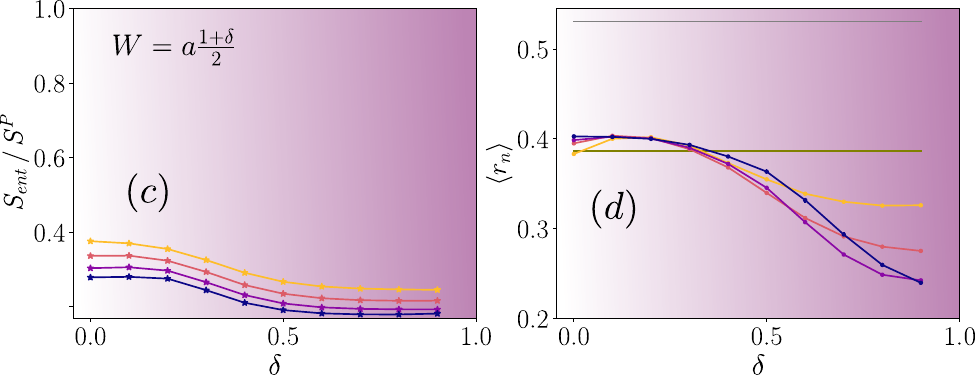}
\caption{Half-chain entanglement entropy $S_{ent}$  and the mean gap ratio $\braket{r_n}$  as a function of the dimerization $\delta$ along the right edge of the parameter space defined via $W=a(1-\delta)/2$ (a-b) and along the left edge,   $W=a(1+\delta)/2$ (c-d).}
    \label{fig:entropy_gap_ratio_edges}
\end{figure}

\begin{figure*}
    \centering
    \includegraphics[width=1.0\linewidth]{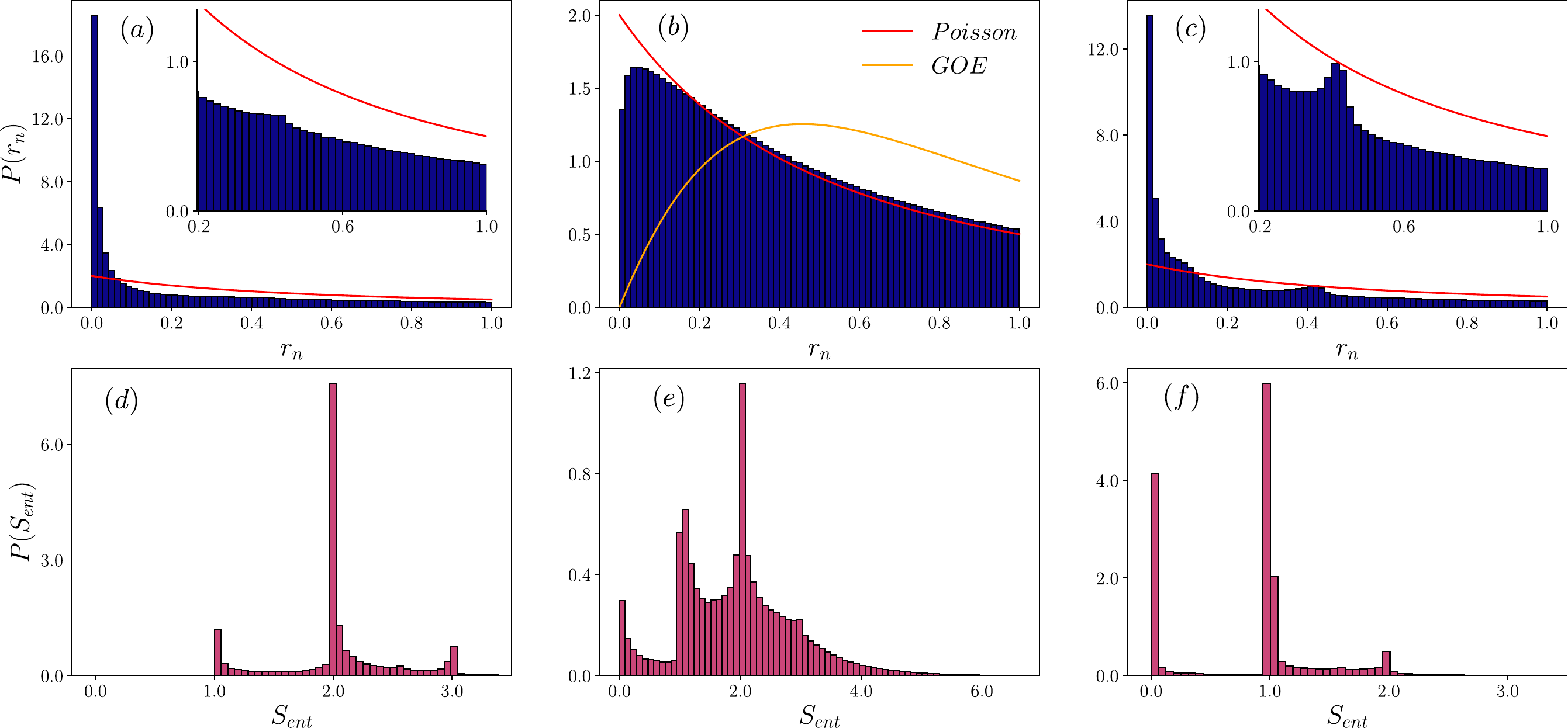}
    
\caption{{Normalized gap ratio distributions (a-c), and distributions of the underlying half-chain entanglement entropies of the eigenstates (d-f).  Plots (a,d) correspond to the system with $W=0.1, \delta = -0.8$, (b,e) - $W=0.5, \delta = 0.0$, and (c,f) - $W=0.1, \delta = 0.8$.  The data correspond to  1000 disorder realizations for $ L = 16$.  Insets in plots (a) and (c) show the zoomed-in tail behavior of the corresponding gap-ratio distributions.}}
    \label{fig:distrib_gap_ratio_entropy}
\end{figure*}

\begin{figure}[ht]
    \centering
    \includegraphics[width=1.0\linewidth]{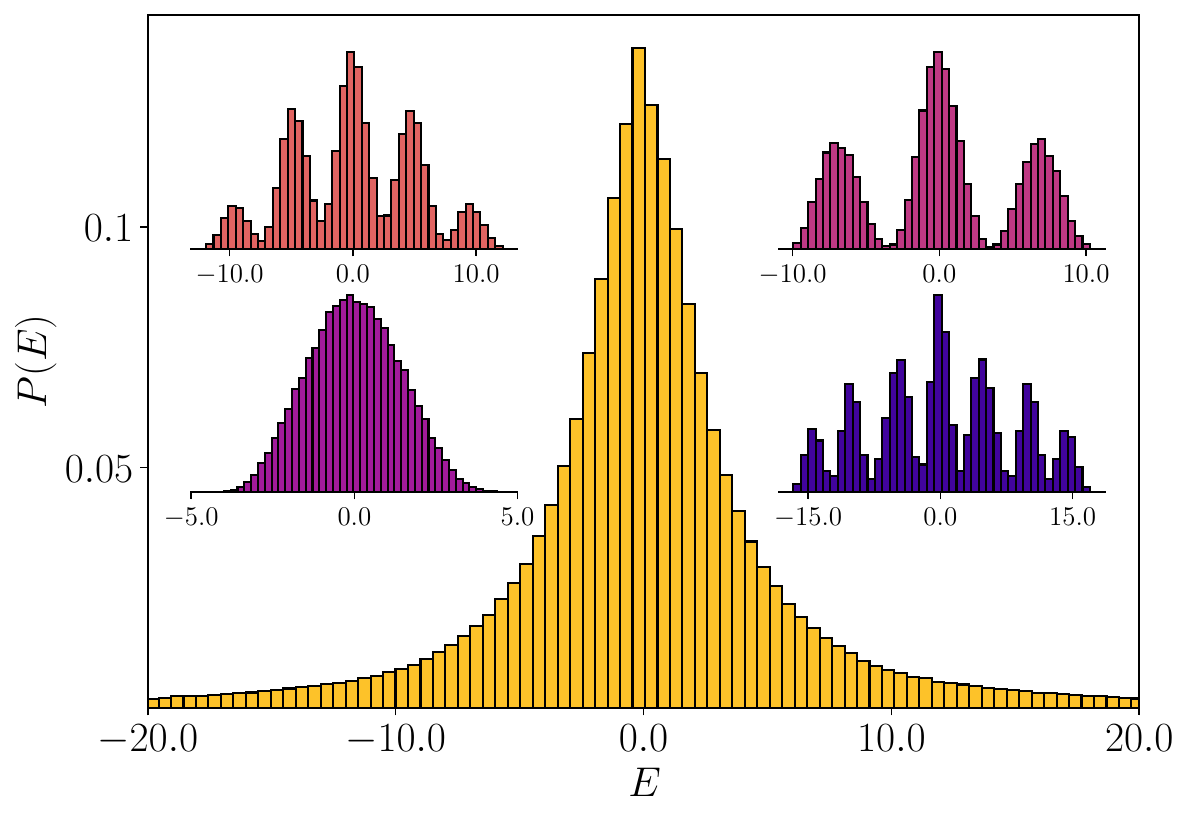}
    \caption{{Normalized histogram representing the density of states obtained by cumulating $1000$ energy spectra corresponding to the different disorder realizations of the model with $ L = 16, W=0.5$ and $\delta=0.0$. The insets show the energy densities of the individual, typical realizations from this ensemble.}}
    \label{fig:distrib_energy}
\end{figure}

In an attempt to further characterize the localized parts of the parameter space,
we consider localization properties of our model along the edges of the triangular parameter space as shown in Fig.~\ref{fig:entropy_gap_ratio_edges}.
The EE shows the sub-volume behavior, and the mean gap ratio remains low, confirming the localized character of eigenstates. Clearly, for the central localized domain (at low $|\delta|$), the mean gap ratio is larger than the Poissonian value regardless of the system size. On the contrary, large $|\delta|$ yields a sub-Poissonian value. No clear crossings of curves for different system sizes are observed, suggesting a smooth crossover between 
localized domains. 

Further insight may be obtained by looking not merely at averages but considering the full distributions as shown in Fig.~\ref{fig:distrib_gap_ratio_entropy} for $L = 16$. The middle column, panels (b, e), corresponds to the large $W$ and small $|\delta|$ case (upper part of the triangle in Fig~\ref{fig:schem}). The system is not fully localized, with a residual small level repulsion as revealed by a dip at small gaps. The EE distribution shows a multimodal structure with prominence of $S_{ent}\gtrsim 1,2$ (note that we use the logarithm with base 2 in the definition of the entanglement entropy). This contrasts with the typical picture in MBL where {$S_{ent}\approx 0$} dominates at strong disorder. Integer values of $S_{ent}$ result either from maximally correlated local subsystems (of 2 sites) or larger frozen domain of spins~\cite{Aramthottil24} forming a cat structure with its spin inversion due to the $\mathbb{Z}_2$ symmetry. Both make a contribution of $S_{ent}=1$ given the correlation or spin domain crosses the subsystem cut for the entanglement entropy.


{The situation is markedly different for both left (negative $\delta$) and right (large positive $\delta$) regions. The level gap ratio shows a characteristic level attraction behaviour. Most of the states have an integer half-chain EE being either one or two (note that the remnants of this structure are also seen in the middle panel). $S_{ent}=2$ appears if, apart from $\mathbb{Z}_2$ symmetry effect, cat states are created involving states on the different sides of the mid-chain cut. The relative density of such states is increased for dimerized chains, which is partially explained by the renormalization scheme discussed below. The presence of such cat states results in a significant amount of very closely spaced level pairs - that explains a prominent excess of small gaps that dominates gap distributions [see top row, panels (a) and (c)]). Interestingly, the gap
distributions reveal a persistent resonant feature around the gap ratio $r_n\approx0.45$, quite pronounced for positive $\delta$. We presently have no explanation for this unusual behavior. It is statistically significant.}

Additional understanding of the localization properties of our model can be gained by looking at the energy densities of the considered Hamiltonian. Fig.~\ref{fig:distrib_energy} shows the normalized histogram containing energies coming from $1000$ realizations of disorder for  $L = 16$ model with $W = 0.5$ and $\delta=0.0$. The distribution deviates from a Gaussian shape, showing instead a Lorentzian-like profile. This shape results from an 
average over different, vastly different individual realizations, some of which 
are depicted as insets in Fig.~\ref{fig:distrib_energy}. First, some realizations yield a Gaussian profile. These typically arise from either small random displacements of individual sites or from collective shifts of all spins that do not induce pairing, such as when all spins are displaced in the same direction. More intricate profiles arise when such spin pairing occurs. The simplest case involves configurations in which most positions experience only mild disorder, remaining close to the undistorted arrangement, except for two spins that are shifted toward each other with a displacement comparable to $W$. Then the bond between the two paired spins becomes dominant, and the structure of the entire many-body spectrum is governed by the four eigenstates of the corresponding local tunneling operator. As a result, the spectrum splits into three distinct islands. Extending this reasoning, one can expect realizations featuring multiple such paired spins distributed along the chain, leading to correspondingly fragmented energy spectra. In this way, we can construct an ensemble of disordered Hamiltonians exhibiting unique Hilbert-space fragmentation patterns determined by the specific disorder configurations.


\section{Spin-glass character of the localized regime}
\label{sec:gl}

{To further delve into the structure of the excited eigenstates we probe the spin glass order \cite{Parisi80,Parisi83}. }To quantify it in our model, we follow \cite{Vasseur16} and apply the Edwards-Anderson spin glass order parameter, which serves as a probe of long-range Ising correlations in  eigenstates and is defined as:
\begin{equation}
    m_{EA} = \frac{1}{L^2} \sum_{n} \sum_{i \neq j} \bra{n} s^z_i s^z_j \ket{n}^2,
   \label{eq:sg} 
\end{equation}
where $\ket{n}$ are eigenstates of the problem {and  $n$ runs over the whole spectrum.} We also consider the rescaled auxiliary quantity 
 $\chi_{EA} = L m_{EA}$ \cite{Kjall14}. Note that our definition \eqref{eq:sg} differs by a factor of 16 from that defined in \cite{Vasseur16} using Pauli matrices.
 
\begin{figure}[ht]
    \centering
    \includegraphics[width=1.0\linewidth]{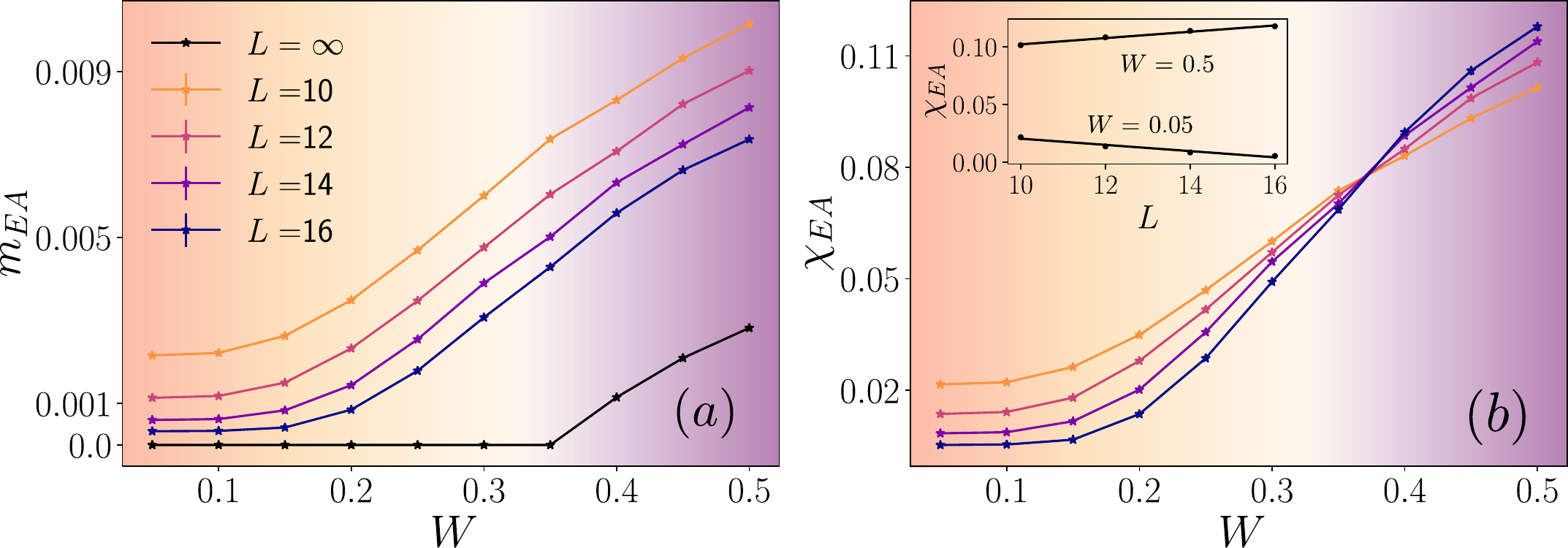}
     \centering
    \includegraphics[width=1.0\linewidth]{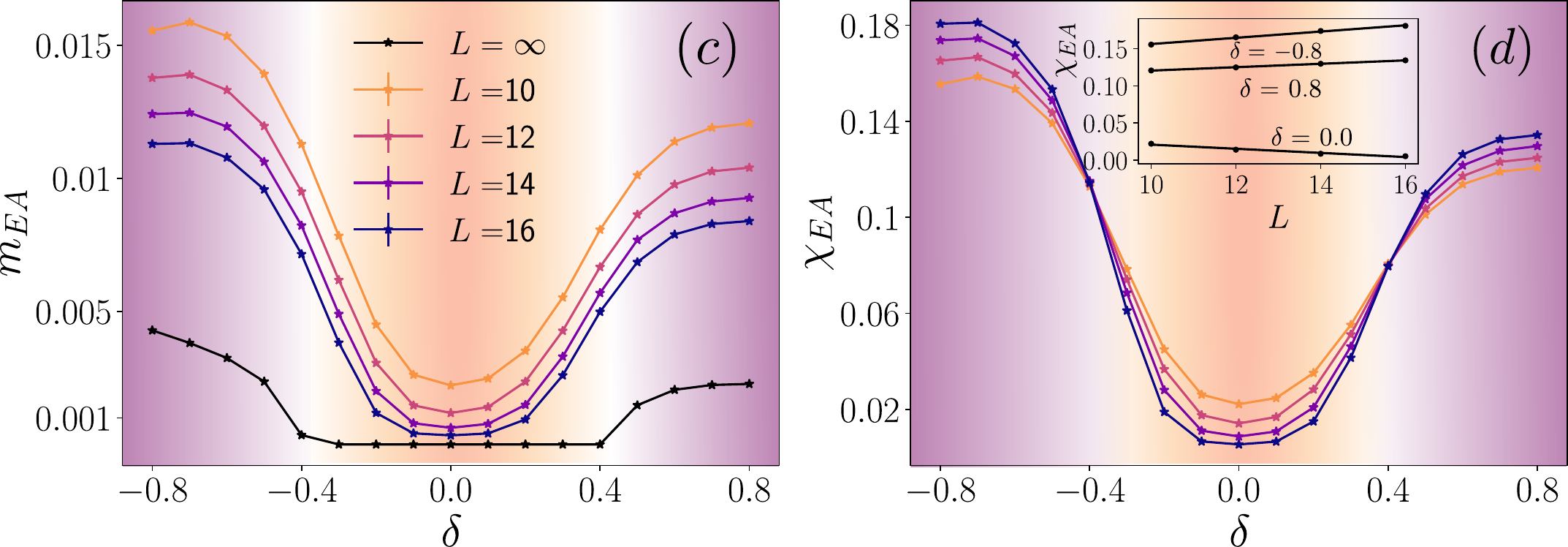}
\caption{Edwards-Anderson order parameter $m_{EA}$  and the auxiliary $\chi_{EA}$ as a function of  the disorder strength $W$ for fixed $\delta = 0$ (a-b) and as a function of dimerization $\delta$ for  fixed $ W= 0.1$ (c-d). Insets on plot (b) and (d) show the scaling of the $\chi_{EA}$ with a system size for particular parameter points.}
    \label{fig:spin_glass}
\end{figure}

The dependence of the spin glass parameter on $W,\delta$ for horizontal and vertical cuts in the parameter space considered in the main text, are shown in  Fig.~\ref{fig:spin_glass}. The change of the disorder amplitude $W$ for $\delta=0$ (vertical cut) leads to the growth of $m_{EA}$ with increasing $W$ and a crossing of the $\chi_{EA}$ curves around $W \approx 0.37$ - c.f. Fig.~\ref{fig:spin_glass}(a,b). Additionally, an extrapolation of  $m_{EA}$ with an inverse of the system size shows a non-vanishing value after the crossover. Similarly, for the horizontal cut [Fig.~\ref{fig:spin_glass}(c,d)] there are two crossovers in $\chi_{EA}$, with two regimes with extrapolated nonzero $m_{EA}$ appearing for sufficiently large $|\delta|$. Insets in panels (b) and (d) of Fig.~\ref{fig:spin_glass} further amplify this observation by showing that $\chi_{EA}$ grows linearly with the system size in the localized regime, while in the ergodic domain it decreases with $L$. 



{
Therefore, the numerical results provide a strong evidence of the spin-glass order hosted by our system in the broken ergodicity phase, even without $ZZ$ interaction terms. For the finite system sizes accessible in our computations, it is evident that only a portion of the spectrum exhibits this behavior; however, in the limit of large $L$, it remains unclear whether the system becomes fully spin glass or whether some fraction of the spectrum retains qualitatively different behavior.

In the next sections we will focus on the strong dimerization limit $|\delta| > 0$ and use RSRG-X predictions and numerical tools to investigate the SPT states across the whole energy range. As discussed in \cite{Vasseur16}, glassy states cannot host SPT order because they spontaneously break the protecting symmetry. Thus, the SPT states must form a distinct subset of eigenstates, and it is interesting to investigate its fate. 
}

\section {Real space renormalization group viewpoint}
\label{RG}

To gain insight into the form of eigenstates in {\it localized regimes}, we use the real space renormalization group for excited states (RSRG-X) \cite{Pekker14}, an extension of the strong disorder renormalization group scheme \cite{Dasgupta80, Fisher94} for excited states. This
scheme is applicable to strong disorder, localized phase only, and works best for Hamiltonians constructed as the sum of operators that act on a few sites with a broad range of energy gaps \cite{Protopopov20,Mohdeb22, 
Aramthottil24}. Our version is specifically adapted to a dimerized Hamiltonian; for a general scheme for long-range interacting systems, see \cite{Zhao25}.

Initially, the two-site operator with the largest coupling is identified in \eqref{eq:model}. To be concrete let us consider for that $i,j=i+1$ sites: $H_{ij}=\frac{J_{ij}}{2}(s_i^+s_{j}^- + s_i^-s_{j}^+)$.  In the resulting dim=4 subspace, the diagonalization gives eigenstates $ \ket{\pm}_{ij} \equiv\frac{1}{\sqrt{2}}(\ket{\uparrow \downarrow} \pm \ket{\downarrow\uparrow})_{ij}$ and a degenerate subspace spanned by $\ket{Z_{+/-}}_{ij}\equiv \ket{\uparrow \uparrow }/ \ket{\downarrow\downarrow}_{ij}$. In this new basis,
one perturbatively modifies operators connected to sites $i,j$, of generic forms $H_{li},H_{lj},H_{ir},H_{jr}$. For $\ket{\pm}_{ij}$ states, the resulting operator  takes the form 
$$H_{lr}=\frac{[J_{li}(J_{jr}\pm J_{ir})+J_{lj}(J_{ir}\pm J_{jr})]}{2J_{ij}}(s_l^+s_r^-+s_l^-s_r^+).$$
The $l,r$ sites are again chosen by the strength of the coupling between them. 

In the degenerate subspace we have an additional possible choice of the basis that affects the functional form of $H_{lr}$. For example, the $\ket{Z_{\pm}}_{ij}$ leads to 
\begin{eqnarray}
H_{lr}&=&-\frac{[J_{li}J_{jr}+J_{lj}J_{ir}]}{2J_{ij}}(s_l^+s_r^-+s_l^-s_r^+)\nonumber\\
&\pm& 2[ \frac{J_{li}J_{lj}}{J_{ij}}s_z^l+\frac{J_{ir}J_{jr}}{J_{ij}}s_z^r]
\nonumber
\end{eqnarray}
yielding chemical potential-like terms while 
$\frac{1}{\sqrt{2}}(\ket{Z_+}_{ij}\pm \ket{Z_-}_{ij})$ results in 
\begin{eqnarray}
H_{lr}&=&-\frac{[J_{li}J_{jr}+J_{lj}J_{ir}]}{2J_{ij}}(s_l^+s_r^-+s_l^-s_r^+) \nonumber\\
&\mp& \frac{[J_{li}J_{ir}+J_{lj}J_{jr}]}{2J_{ij}}(s_l^+s_r^++s_l^-s_r^-) \nonumber
\end{eqnarray}
generating  a pair-hopping term. The chemical potential term effectively creates domains of parallel spins (responsible for spin-glass order), the pair-hopping term creates $Z-$exchanges: $\ket{Z_{\pm}}_{ij} \rightleftarrows \ket{Z_{\mp}}_{lr}$. We follow this last choice if the energy gap associated with the exchange is greater than the gap resulting from a domain with the least distance to the pair. 

\begin{figure}[ht]
    \centering
    \includegraphics[width=1.0\linewidth]{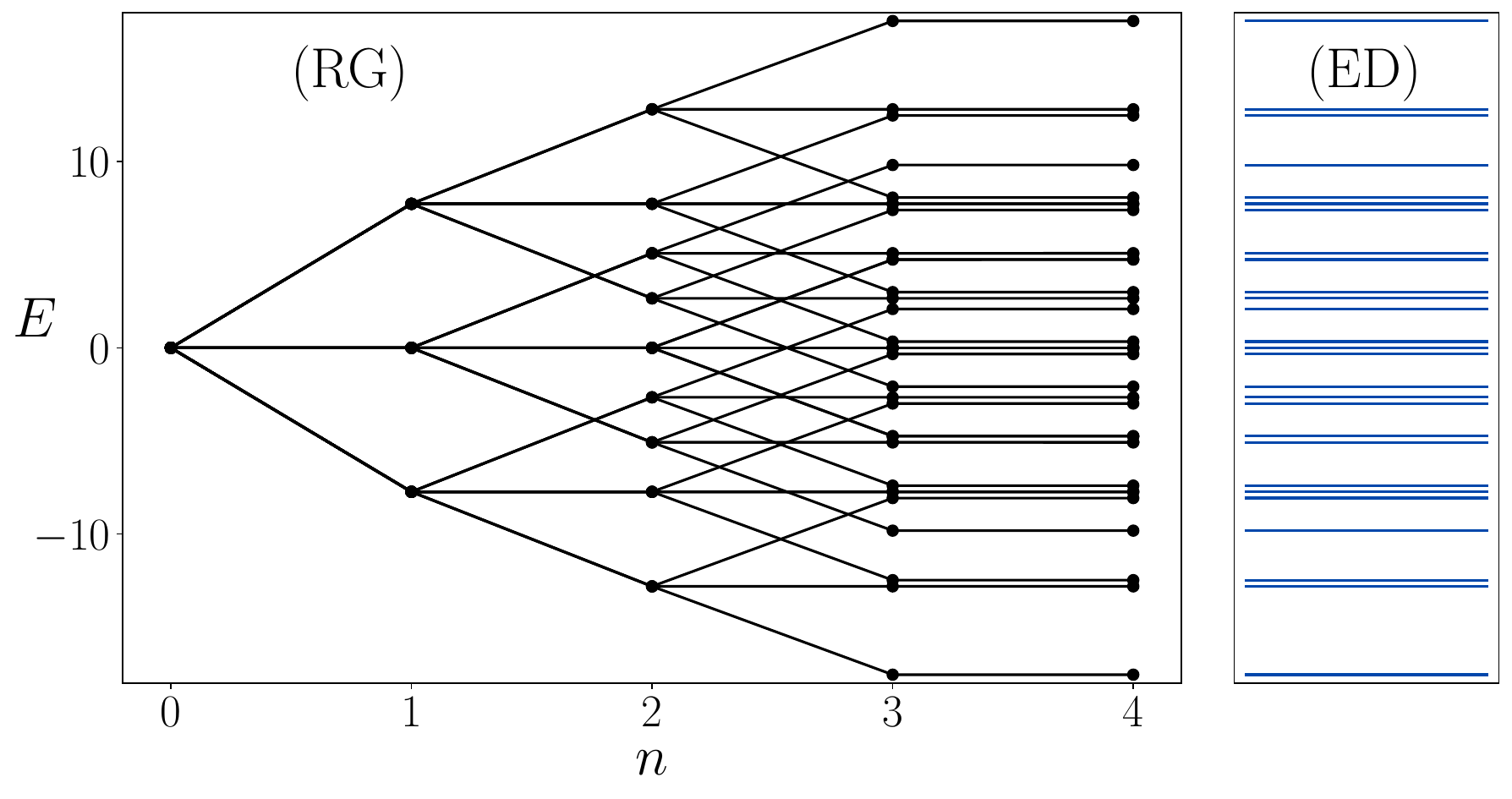}    
\caption{Comparison of energy levels resulting from the RG procedure for $L=8$ chain with results of the exact diagionalization shown in the right panel. In the left panel consecutive steps $n$ of the RG scheme are depicted up to $n=4$ when the spectrum is resolved. The data correspond to $\delta =-0.8$ and $W=0.1$.}
    \label{fig:comp}
\end{figure}
In this way, four sites $l,i,j,r$ are replaced by two sites $l,r$ with effective operator $H_{lr}$. The procedure is called a decimation. Successive decimations involve more sites. Spanning all the chain and diagonalizing the last operator gives the estimation of eigenstates for a given path ${\cal P}$, defined by successive choices of basis states during decimations.

In the positive $\delta$ regime, the bonds between sites $(2i-1,2i)$ will be the strong bonds leading to large energy gaps. Thus, the operators that are fixed in decimations will be on the sites occupied by these bonds. Depending on the path ${\cal P}$ chosen, either via cat states $\ket{\pm}$ or $Z$-subspace, this results in approximate eigenstates of the form $\ket{\Psi_\mathrm{T}} = \otimes_{\lbrace j\rbrace} \ket{\pm}_{2j-1,2j} \otimes_{\lbrace i \rbrace}( \cos(\varphi_i)\ket{Z_+}_{2i-1,2i}+\sin(\varphi_i)\ket{Z_-}_{2i-1,2i})$ where, $\lbrace j\rbrace $ are sites involving $\ket{\pm}$ and $\lbrace i\rbrace $ involving linear combinations (defined via $\varphi$) of $\ket{Z_\pm}$ states.

In contrast, in the $\delta<0$ regime, the bonds between sites $(2i,2i+1)$ will be comparatively larger and produce the largest energy gaps. This leads to isolated spins at the boundaries that are connected to the rest by weak bonds only. Those will be coupled in the last decimation step.  Once all relevant bonds in the bulk are fixed, the most relevant bonds associated with the edge sites will be the bond that directly connects them, of the energy scale $J\approx 1/(L\tilde{a})^3$, or the chemical potential bond from the nearby dimer, which will go as $J\approx [\delta a+\tilde{a}]^3/(m\tilde{a})^6$, where $m\tilde{a}$ is the distance to the nearby fixed $\ket{Z_{\pm}}$ domain.  The cases of states where the direct connection dominates can have edge sites fixed as $\ket{\pm}_{1,L}$ or $\ket{Z_{\pm}}_{1,L}$. In the latter case, one can have exchanges with the bulk. We shall refer to both such types of states as $\ket{\Psi_{\mathrm{NT}}}$.

 As a simple illustrative example of the RG steps, we consider an $L=8$ chain in zero magnetization sector. We progressively fix the two sites with the largest energy gap with it's eigenstates and append the energies (the perturbative corrections are not considered here) referred to as a decimation of step, $n$. The energies found after fixing the last pair of sites are found to faithfully reproduce the energies obtained from exact diagonalization as 
shown in Fig.~\ref{fig:comp}. 

Our notation of $\ket{\Psi_\mathrm{T}}$ and $\ket{\Psi_{\mathrm{NT}}}$ states is intentional. We shall argue in the next section that those states have, respectively, trivial; and non-trivial topological character.

\section{Implications for the SPT order}
\label{sec:SPT}

Distinguishing between different SPT phases numerically is a challenging task \cite{Pollmann12}. The often-used degeneracy of the entanglement spectrum fails to detect the topological character in the presence of cat states \cite{Pollmann10}. Indeed, this is the case for our model.
{
The standard EE shows the 4-fold degeneracies of the entanglement spectrum in the localized regime for appropriate entanglement cuts, that may come from cat-like states in the bulk or involving the edges of the chain.}

Recently, it was shown \cite{Zeng16, Zengbook}, however,  that the long-range entanglement between edge states 
can identify nontrivial SPT order. 
That can be quantified by the so-called disconnected entropy \cite{Zeng16, Fromholz20}:
\begin{equation}
    \label{eq:disco}
    S^D =S^{A}_{ent}+S^{B}_{ent}-S^{A \cup B}_{ent}-S^{A \cap B}_{ent},
\end{equation}
where $S_{ent}^{C}$ is a bipartite entanglement entropy for partition $C$. 

\begin{figure}
    \centering
    \includegraphics[width=0.9\linewidth]{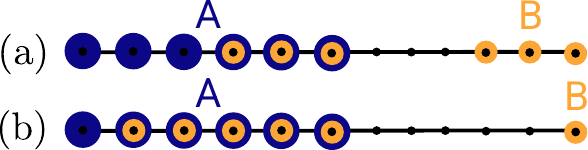}

   \caption{{ Partitions used to define the disconnected entropy (\ref{eq:disco}). Blue and orange circles denote sets $A$ and $B$, respectively. (a) - used for $S^D_1$ and (b) - for $S^D_2$.} }
    \label{fig:podzial}
\end{figure}

Typically, the construction of subsystems is done by a division of the chain into 4 parts,
\cite{Zeng16, Zengbook, Fromholz20,  Micallo20} 
or more formally one can define $A =\left\{1, ... L/2\right\}$ and  $B = \left\{  \lceil L/4  \rceil+1, ... L/2,  \lfloor 3 L / 4  \rfloor + 1, ... L\right\}$ -- c.f. Fig.\ref{fig:podzial}(a) -- where floor $\lfloor  x\rfloor = \mathrm{max} \{z \in\mathbb{Z} | z \le x \}$ and ceiling $\lceil  x\rceil = \mathrm{min} \{z \in\mathbb{Z} | z \ge x \}$ are introduced to deal with system sizes which are not divisible by 4. For reasons, which will be apparent below, we propose another way to divide our chain: 
$A =\left\{1, ... L/2 + (L/2 \bmod 2)\right\}$,  $B = \left\{  2, ... L/2 + (L/2 \bmod  2), L\right\}$ as shown in  Fig.\ref{fig:podzial}(b). Let us denote disconnected entropies corresponding to these conventions as $S^{D}_{1}$ for the original one and $S^{D}_{2}$ for our definition. 

As a first application, let us consider the ground state properties.
The Hamiltonian (\ref{eq:model}) is an extension of the bosonic version of the SSH model \cite{Su79,Hasan10,Kane13,
Batra20} by long-range hopping (that preserves the symmetries of the system). 
\begin{figure}
    \centering
    \includegraphics[width=0.9\linewidth]{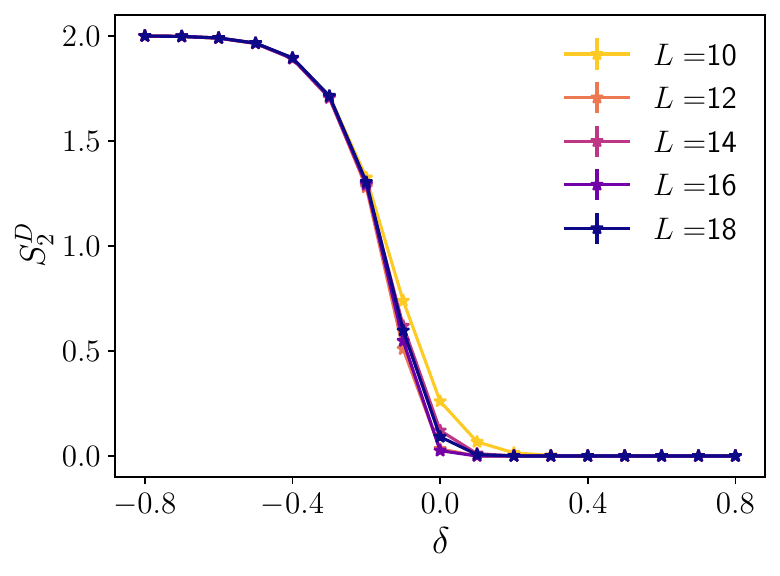}
    \caption{Average disconnected entropy $S^D_2$ for the ground state of \eqref{eq:model} for disorder amplitude $W=0.1$ as a function of $\delta$ for different system sizes.}
    \label{fig:groundstate}
\end{figure}
Its ground states, for negative/positive $\delta$, even in the presence of the positional disorder, are in nontrivial/trivial SPT phases - c.f. Fig.~\ref{fig:groundstate}. For $\delta < 0$, in the nontrivial phase, the ground state manifold is characterized by the presence of edge modes (with the corresponding 4-fold degeneracy in the open chain) and a long-range string order \cite{Torre06, Berg08}. Additionally, these states can not be smoothly deformed into ground states coming from the trivial phase ($\delta > 0$), without breaking the protecting symmetry or breaking the condition of short-range entanglement (e.g by closing the gap). The corresponding disconnected entropy $S^D_{1,2}$ saturates to $2$. On the other hand, for positive $\delta$ the ground state is trivial topologically, with $S^D_{1,2}$ approaching $0$. Therefore, nontrivial short-entangled states  
are characterized by nonzero quantized values of the disconnected entropies, $S^D_{1,2}$.

Let us now consider the excited states.
The RSRG-X procedure, described above,  strongly suggests that our model in the localized regimes hosts a significant fraction of short-correlated states of different topologies across the entire excited spectrum. 

\begin{figure}
    \centering
    \includegraphics[width=1.0\linewidth]{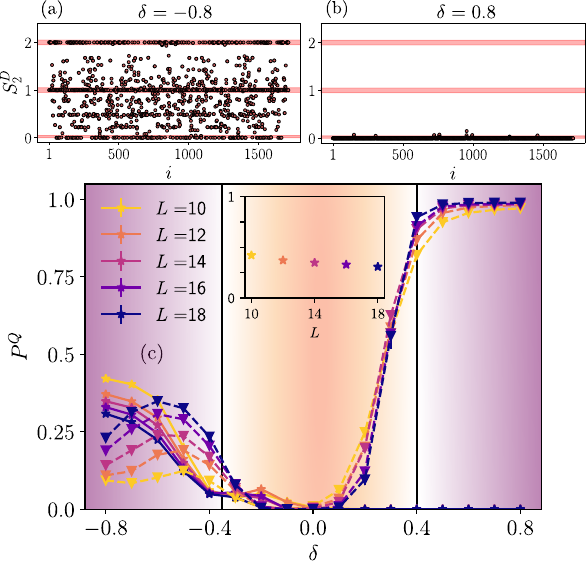}
    \caption{ Panels (a-b): $S^{D}_{2}$ for all states, parametrized by their eigen index $i$ (increasing with energy),  within the appropriate symmetry sector at $L=14$, for a single typical disorder realization for $W=0.1$ with $ \delta = -0.8$ and $\delta = 0.8$, respectively. (c) Fraction of states with $S^{D}_{2} =1,2$  is indicated by stars connected by solid line for different system sizes. Triangles connected by dashed lines represent the fraction of states with $S^{D}_{2} = 0$. Data averaged over several disorder realizations for $W=0.1$. The inset shows the fraction of nontrivial states for $\delta=-0.8$.}
    \label{fig:disco2}
\end{figure}

 Consider first the sufficiently negative $\delta$. We expect to have $\ket{\Psi_{\mathrm{NT}}}$ eigenstates across the spectrum. Firstly if we fix  $\ket{\phi}_{1,L} = \ket{\pm}$, these states will have the same construction as the nontrivial groundstate, and more complicated structures will be coming from bulk exchanges, which are analogous to triplon excitations studied in \cite{Chandran14}. 
Such states will have $S^{D}_{1} = S^{D}_{2} = 2$ indicating maximal correlations between edge sites. They can be interpreted as topologically nontrivial.
The next simple family for the same parameters consists of the states $\ket{\Psi_{\mathrm{NT}}}$ with $\ket{\phi}_{1,L} = \ket{Z_{\pm}}$.
{Those $\ket{Z_{\pm}}$ domains get maximally correlated with a pair of sites in the bulk $\ket{\phi}_{l,r}=\ket{Z_{\mp}}$ (the choice of the sign is dictated by the symmetry sector considered). Such $\ket{\Psi_{\mathrm{NT}}}$
states can be interpreted as domain walls in the bulk of the nontrivial SSH chain.}
These states have a quantized value of $S^{D}_{2} = 1$, meaning that edge spins are coherently correlated, but entanglement is not maximal. For the original definition of $S^{D}_{1}$, it is more ambiguous, because we are looking at correlations between multiple sites close to the edges. If a domain wall is formed in these areas, $S^{D}_{1}$ will give $2$, hence it does not distinguish between these states and states from the first family. 

{Let us note that we have verified, in chosen cases, that states with $S^D_2=2$ are 4-fold degenerate in energy when all possible symmetry sectors are considered. This further strengthens the interpretation that those states are topologically nontrivial. On the other hand, such a check is numerically costly as it requires separate diagonalizations in different energy sectors.}

 The numerical results for $S^D_2$ are shown in Fig.~\ref{fig:disco2}(a,c) (results for $S^D_1$ are shown in the Appendix for comparison). We plot the fraction of states with quantized, integer values of disconnected entropy
\begin{equation}
    P^{Q}(S^{D} = k) = \left<{N_{k}}/{N}\right>,
    \label{eq:P_Q}
\end{equation}
where $N_k$ is the number of the states with $S^D \in [k - \epsilon, k + \epsilon]$, $k = 0, 1, 2$, $N$ - the dimension of the corresponding symmetry block and averaging $\left< \cdot \right>$ is taken over disorder realizations. We set $\epsilon=0.05$ in all of our calculations.
One may observe the presence of a significant number of states with $S^{D}_{2} = 0$, suggesting that not all "product states" in this configuration of spins will develop coherent correlations between edges. In particular, states with non-integer values of $S^D$ have a more complicated entanglement structure and, typically, larger entanglement entropy. On the other hand, a significant fraction of states has non-zero, quantized $S^D_2$, indicating their topological character. 
To estimate this fraction, full spectra of \eqref{eq:model} were needed, so $50$ disorder realizations were used for $L=18$. 

\begin{figure}
    \centering
    \includegraphics[width=1.0\linewidth]{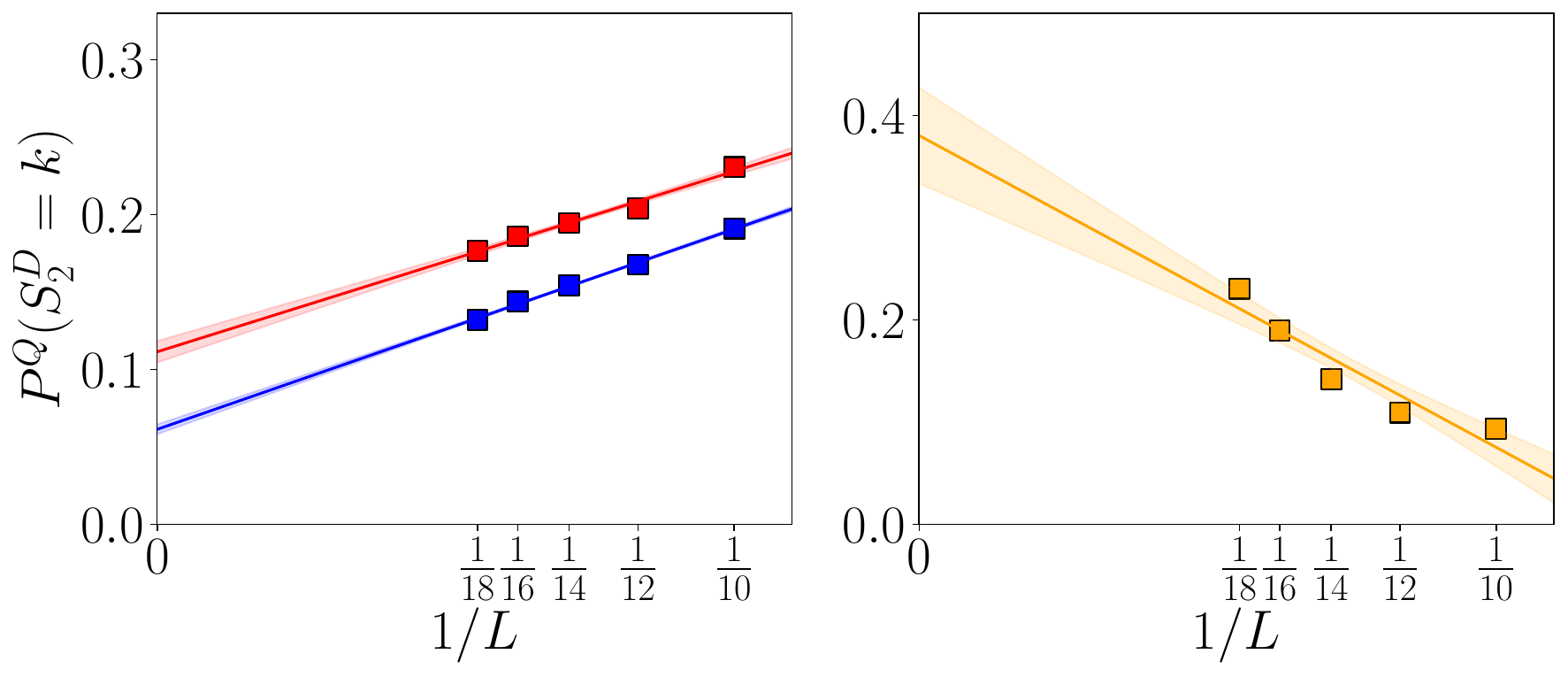}
 
    \caption{{Scaling of the $P^Q(S^D_2)$ values for $W=0.1$ and $\delta=-0.8$ with the inverse of the system size. The squares correspond to the disorder averages of the numerical data. Fitted curves and underlying data of the blue color correspond to the $k = 2$, red color - $k = 1$ and orange - $k = 0$. }}
    \label{fig:scaling}
\end{figure}

One can wonder whether a significant fraction of states with integer values of $S^D_2$ survives in the thermodynamic limit. While we cannot provide a definite answer based on our numerical data, their extrapolation to the infinite size limit, as shown in Fig.~\ref{fig:scaling} is quite suggestive. Thus, despite the finite-size behavior of the Edwards–Anderson order parameter, the system may retain a significant fraction of SPT states in the limit of large $L$ across the whole energy spectrum. {To clarify the relation between nontrivial SPT states and spin-glass order, we observe that these states exhibit significantly weaker $ZZ$ correlations than the rest of the spectrum. As a result, their contribution to the Edwards–Anderson order parameter is nonsignificant. Consequently, even if an extensive number of these states survives in the thermodynamic limit, we expect the spin-glass order to be driven primarily by the complementary part of the spectrum.}

For sufficiently positive $\delta$ we expect formation of {trivial topologically} states $\ket{\Psi_{\mathrm{T}}}$. 
Again, deviations from the simple dimer product state structure are captured by increasingly complex exchanges of frozen spins on our dimers. $S^{D}_{2}$ will give $0$ for this type of states, because arrangement of coherent entanglement between the first and last spins is impossible within this configuration. Strikingly, this picture is different for $S^{D}_{1}$. Frozen spin exchanges can happen at the ends of our chain, and this will lead to $S^{D}_{1} = 2$. Similar states with $S^{D}_{1} = 1$ also may occur. For that reason, we defined $S^D_2$ as a more informative disconnected entropy in our case.

Let us just mention that in the ergodic regime ($\delta$ around $0$) we expect to have arbitrary (unquantized) values of the disconnected entropy, because eigenstates across the spectrum are volume-law entangled.

\begin{figure}
    \centering
    \includegraphics[width=1.0\linewidth]{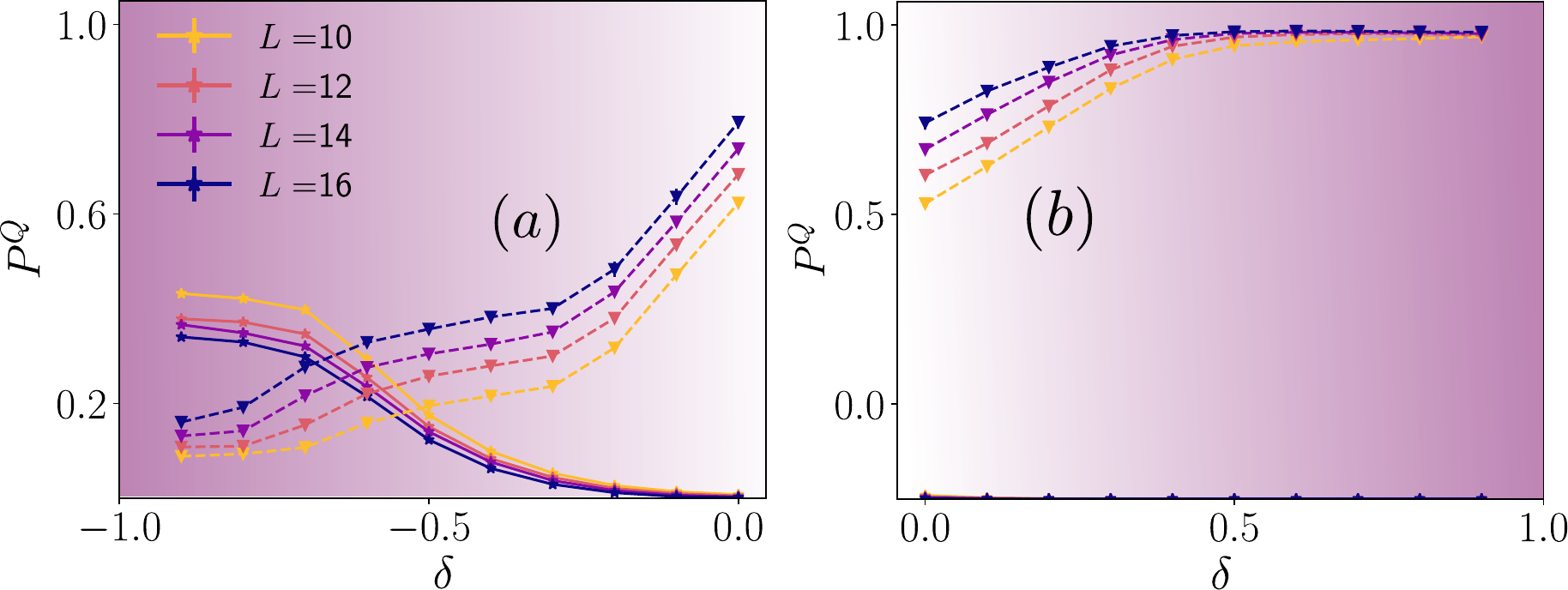}
    \caption{Fractions of states characterized by integer $S^D_2$ values along the left edge of the triangular parameter space (left) i.e. for $W=(1+\delta)/2$ for negative $\delta$ and along the right edge  for $W=(1-\delta)/2$ (right). In the latter case we observe a significant fraction of $S^D_2=0$ states only as indicated by triangles. On the other hand, for a strongly negative delta, close to the lower left corner of the triangle in Fig.~\ref{fig:schem} the fraction of nontrivial states with $S^D_2=1$ or $S^D_2=2$ (as indicated by stars) is significant. }
    \label{fig:disco2_egde}
\end{figure}
To complete the picture, we show in Fig.~\ref{fig:disco2_egde}  how the fractions of states with integer-valued $S^D_2$ change when both the disorder and dimerization change along the edges of the triangular parameter space (compare Fig.~\ref{fig:schem}). Close to the lower left corner, in the strongly dimerized regime, the nontrivial states, i.e., states with integer, non-zero $S^D_2$ are quite abundant. Their number smoothly decreases when dimerization decreases and disorder increases. On the contrary, the number of trivial states increases, those states dominate the spectrum for large disorder or, even more, for a positive dimerization, $\delta$. 

\section{Time dynamics}
\label{sec:time}
\begin{figure}
    \centering
    \includegraphics[width=\linewidth]{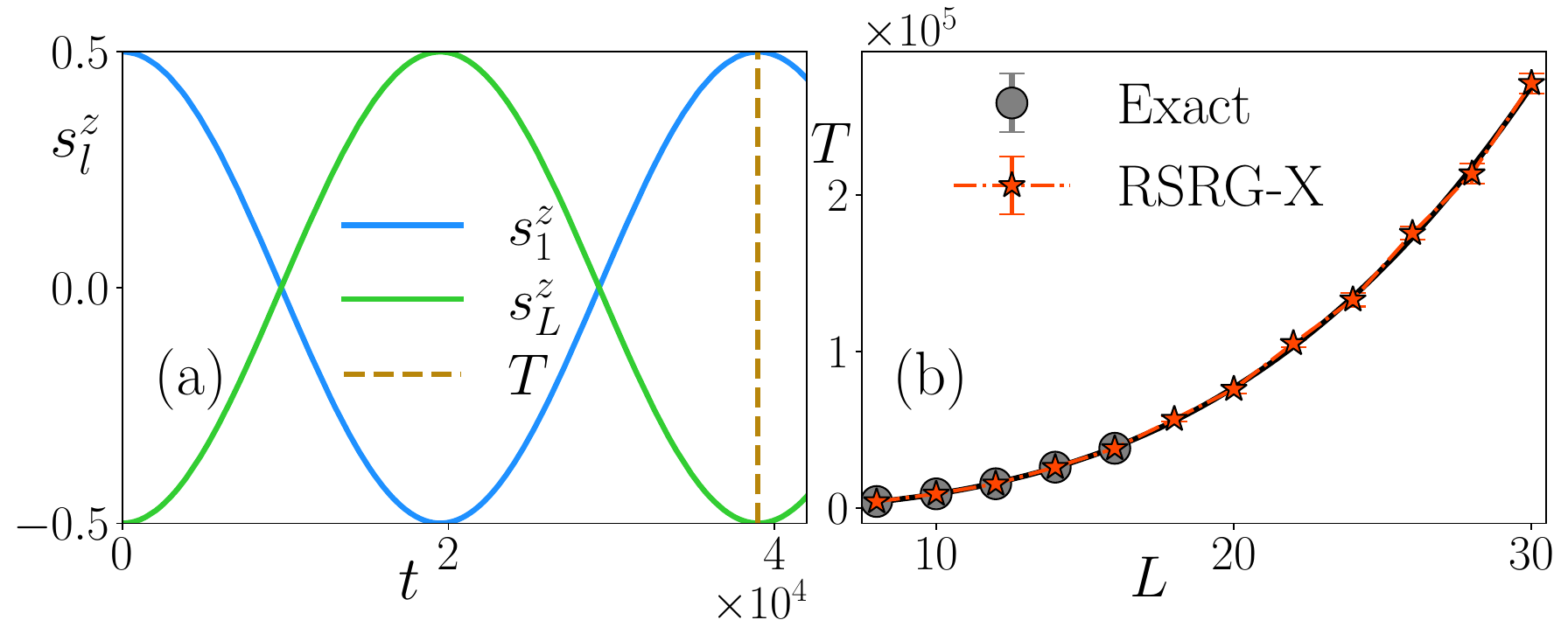}
    \caption{(a) Represents oscillations observed in site-resolved magnetization, $s_l^z$, of the first and last sites for a single disorder realization for the quench state, $\ket{\Psi}$, in the nontrivial regime ($L=16$). The dashed line indicates the time period, $T$, predicted by RSRG-X. (b) Disorder-averaged values{, with a minimum of 200 realizations,} for such $T$s obtained from both exact and RSRG-X for various system sizes are plotted. The black curve represents a third-order polynomial fit. The system parameters for both plots are $\delta=-0.8$ and $W=0.1$.
}
    \label{fig:N_C}
\end{figure}

It is interesting to note, that the presence of excited topological states may be verified by a study of time dynamics from well prepared initial states. This selection is made easy by the RG picture.
Consider the time evolution of an initial state with alternate dimerizations of $\ket{\pm}$ states in the bulk, and frozen opposite spins at the edge sites:  $\ket{\Psi}\equiv  \ket{\uparrow }_1 \ket{+}_{2,3}\ket{-}_{4,5}\cdots \ket{\downarrow}_L$. We take care that the state falls into the middle of the spectrum. In the strong disorder limit, this state can be seen as a linear combination of two approximate eigenstates suggested by RSRG-X: $\ket{\Psi}\equiv \frac{1}{\sqrt{2}}(\ket{\Psi_1}+\ket{\Psi_2})$, of the form $\ket{\Psi}_{1/2}\equiv  \ket{\pm }_{1,L} \ket{+}_{2,3}\ket{-}_{4,5}\cdots \ket{-}_{L-2,L-1}$, with respective energy $E_1$ and $E_2$. As time evolves, the site-resolved magnetization at the edges, ${s}_{1}^z,{s}_{L}^z$  indeed reveal oscillations with a period $T = \frac{2\pi}{E_1-E_2}$ as shown in Fig.~\ref{fig:N_C}(a) for a single disorder realization. 
Figure \ref{fig:N_C}(b) confirms that the disorder-averaged period of oscillations scales as a third power of the system size, as expected from the tunneling part of the Hamiltonian \eqref{eq:model}. 

\begin{figure}
    \centering
    \includegraphics[width=\linewidth]{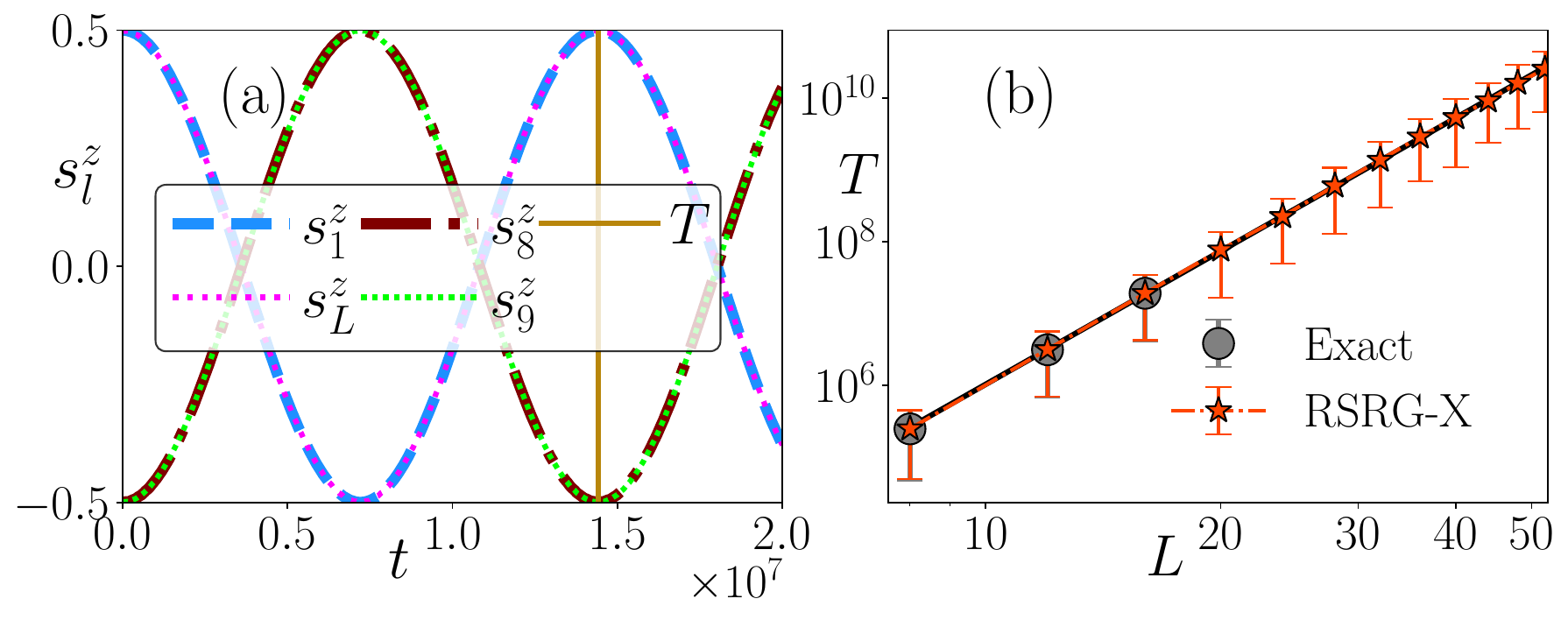}
    \caption{(a) Oscillations observed in site-resolved magnetizations, $s_l^z$, of the first and last sites for a single disorder realization for the quench state, $\ket{\Psi}$, in the nontrivial regime ($L=16$). {Similar oscillations appear for spins located in the middle of the chain.}
    The bold line indicates the period, $T$, predicted by RSRG-X. (b) Disorder-averaged values{, with a minimum of 100 realizations,} for the oscillation period obtained from both exact evolution and RSRG-X for various system sizes are plotted in a log-log scale. The straight line fit yields $\ln(T)=6.17 \ln(L)-0.35$. The system parameters for both plots are $\delta=-0.8$ and $W=0.1$.
}
    \label{fig:Z_E}
\end{figure}

 As another manifestation of the accuracy of the physical picture obtained using the RSRG-X scheme, let us consider the
time dynamics of an initial state of the form $\Psi \equiv \ket{\uparrow}_1 \otimes \ket{-}_{2,3} \otimes \cdots \otimes\ket{Z_-}_{\frac{L}{2},\frac{L}{2}+1} \otimes \ket{+}_{\frac{L}{2}+2,\frac{L}{2}+3} \cdots \otimes \ket{\uparrow}_L$. This state, with energy lying in the middle of the spectrum, may be considered as a superposition of two $\ket{Z}$ edge-bulk exchange states. Looking at the time-evolution of this state one finds the oscillations between the edge $\ket{Z_+}_{1,L}$ and the $\ket{Z_-}_{\frac{L}{2},\frac{L}{2}+1}$ as shown for a particular disorder realization in Fig.~\ref{fig:Z_E}(a). Figure \ref{fig:Z_E}(b) presents the average period, $T$,  found after disorder averaging. As the leading energy scale difference between the eigenstates that form such a state is just $\frac{J_{1\frac{L}{2}+1}J_{\frac{L}{2}+1L}}{J_{\frac{L}{2}\frac{L}{2}+1}}$, the associated period, $T$, is expected to grow as $\propto L^6$ as verified by the fit obtained in Fig.~\ref{fig:Z_E} (b).

\section{Conclusions}
\label{sec:conc}

We have explored the interplay between positional dimerization and disorder in a simple long-range XY open chain, relevant to experiments with Rydberg atoms in optical tweezers \cite{Browaeys20}. We investigated how these effects influence the model’s localization properties, showing that the system does not exhibit standard many-body localization \cite{Nandkishore15, Abanin19}, but instead displays an unusual form of Hilbert space fragmentation \cite{Rakovszky20, Khemani20} driven by the spatial arrangement of spins. Interestingly, our system develops the spin glass order (for a part of the states), despite not having the $ZZ$ interactions. Since dimerization represents one of the simplest routes into topological physics, it further allowed us to examine the interplay between localization and topology. Backed by the strong disorder real space renormalization group arguments \cite{Pekker14}, we discussed the challenges of numerically identifying nontrivial excited SPT states in such open chains and how these can be addressed using modern approaches such as entanglement measures based on disconnected partitions. Our numerical results suggest that the fraction of SPT states may survive in the thermodynamic limit.  Finally, we analyzed the many-body dynamics of the fine-tuned states, showing that it is governed by a few nontrivial SPT eigenstates located in the middle of the spectrum.

 Let us mention that adding additional $ZZ$ interactions to the model, such as that resulting i.e from van der Walls interactions \cite{Homeier25,Zeybek23}, significantly changes its properties, making non-trivial states less abundant as shown in the Appendix~\ref{appC}.

After our submission to arXiv, we became aware of a parallel experimental work \cite{yue25}
where a structural disorder was applied in studies of average topological properties in a spin chain.

\acknowledgments 
 The work of M.P. and J.Z.  was funded by the National Science Centre, Poland, under the OPUS call within the WEAVE program 2021/43/I/ST3/01142.
 A support by the Strategic Programme Excellence Initiative (DIGIWorld) at Jagiellonian University is also acknowledged. We gratefully acknowledge the Polish high-performance computing infrastructure PLGrid (HPC Centers: ACK Cyfronet AGH) for providing computer facilities and support within the computational grant no. PLG/2025/018400.
 
\section*{Data Availability}
 The data that support the findings of this article are openly available at \cite{data1}.

%

\appendix

\section{A different choice of the disconnected entropy}
\label{appA}

We present here the results obtained using a more common definition of the disconnected entropy, $S^D_1$ as defined in \cite{Zeng16, Zengbook, Fromholz20,  Micallo20}. We consider the fraction of states with quantized, integer values of disconnected entropies $P^{Q}$. Recall that integer values appear because we define the entropy with base-2 logarithm. We plot
\begin{equation}
    P^{Q}(S^{D} = k) = \left<{N_{k}}/{N}\right>,
    \label{eq:P_Qa}
\end{equation}
where $N_k$ is the number of the states with $S^D \in [k - \epsilon, k + \epsilon]$, $k = 0, 1, 2$, $N$ - the dimension of the corresponding symmetry sector and averaging $\left< \cdot \right>$ is taken over disorder realizations. We set $\epsilon = 0.05$ in all of our calculations.

\begin{figure}
    \centering
    \includegraphics[width=0.9\linewidth]{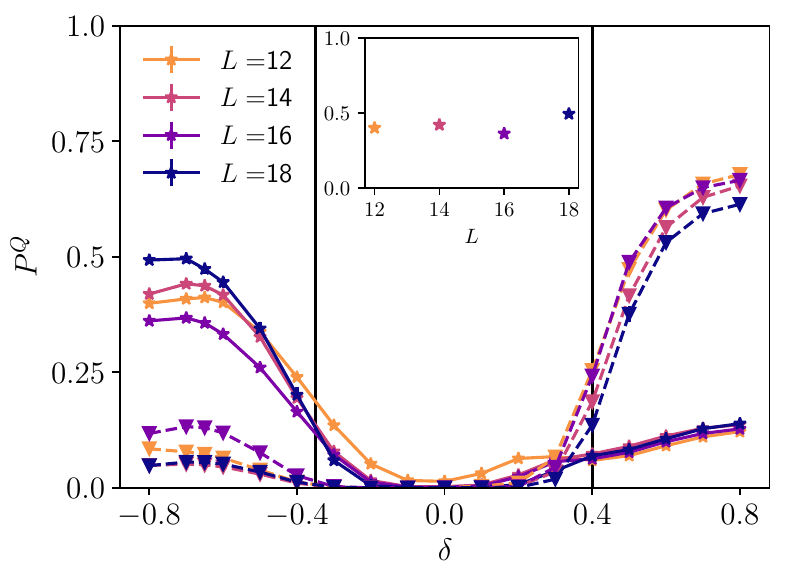}
    \caption{ Fraction of states with $S^{D}_{1} =1,2$  is indicated by stars for different system sizes. Triangles connected by dashed lines represent a fraction of states with $S^{D}_{1} = 0$. Data averaged over several disorder realizations for $W=0.1$. The inset shows the fraction of states with  $S^{D}_{1} =1,2$ for $\delta=-0.8$, showing weak system size dependence.}
    \label{fig:disco1}
\end{figure}

Regions of nonzero $P^{Q}$ are vastly different for $S^D_1$ and $S^{D}_2$. $S^D_1$ states with $k =1, 2$ populate almost whole parameter space (except the ergodic region), since $S^D_1$ captures correlations between bigger partitions around the edges (recall its definition in the main text). Not all of these long-range correlations have a topological nature, because they do not necessarily involve decoupled edge spins. Nevertheless, $S^D_1$ reveals the long-range entanglement structure of the excited eigenstates in our model. We plot $P^{Q}(S^D_1)$ for different system sizes for horizontal cut $W=0.1$ in Fig.\ref{fig:disco1} (in the similar fashion to  Fig.~\ref{fig:disco2}).

For $S^D_2$, the corresponding region with a significant fraction of states with $k =1, 2$ is smaller, such states mainly concentrate in the left bottom corner of the parameter space. It is worth noting, that $S^D_2$ also shows quantized behavior for the nondisordered case $W=0.0$ due to the restoration of the reflection (around the middle of the chain) symmetry, which creates long-range entanglement picked up by $S^D_2$. The corresponding eigenstates do not have the same properties and simple structure as previously discussed ones coming from disordered Hamiltonian (1), so this false signature can be regarded as a drawback of $S^D_2$.

\section{Possible extensions of the model}
\label{appC}

\begin{figure}[h]
    \centering
    \includegraphics[width=1.0\linewidth]{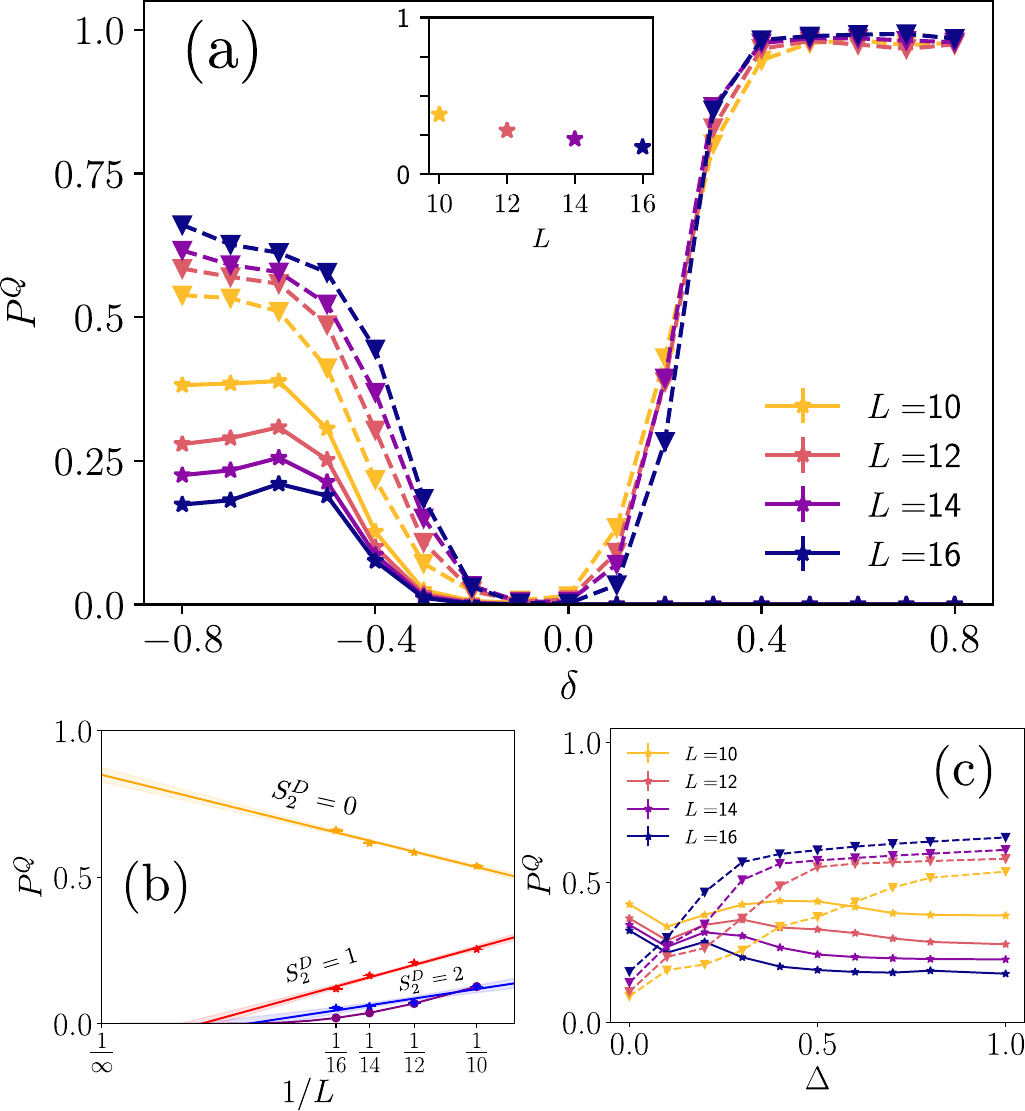}

    \caption{(a) Assuming $\Delta = 1.0$ in (\ref{eq:model1}) fraction of states with $S^{D}_{2} =1,2$  is indicated by stars for different system sizes. Triangles connected by dashed lines represent a fraction of states with $S^{D}_{1} = 0$. Data averaged over several disorder realizations for $W=0.1$. The inset shows the fraction of states with  $S^{D}_{2} =1,2$ for $\delta=-0.8$. (b) Scaling of $P^Q(S^D_2)$ with $\frac{1}{L}$ for $\delta = -0.8$. Purple points show fraction of states without bulk exchanges. (c) Dependence of $P^Q$ on $\Delta$ for $W = 0.1$ and $\delta=-0.8$.}
    \label{fig:disco_nonzero_delta}
\end{figure}

As mentioned in the main text a typical examples of Rydberg atom tweezer arrays without disorder are often well approximated by the pure XY model \cite{Browaeys20,Chen23}. Arranging for the positional disorder is then easy by rearrangement of tweezers. It is interesting, however, to consider 
an enlarged family of models 
 turning on $ZZ$ coupling. Already in \cite{Vasseur16} it was shown (based on RG considerations and numerical results)
 that inclusion of $ZZ$ couplings that decay in the same fashion as tunnelings (in our case $\propto \frac{1}{r^3}$) will destroy all the interesting features at the finite energy density. Following the recent approaches of \cite{Homeier25} and \cite{Zeybek23}, we also check whether the same holds for models that combine tunnelings that scale as $\propto 1 / r_{ij}^3$ arising from direct dipole-dipole interactions between Rydberg states of different parity and van-der-Waals interaction terms scaling as $ \propto 1 / r_{ij}^6$. Consider the  Hamiltonian:

\begin{equation}
    \label{eq:model1}
         H = J\sum_{i>j} \frac{1}{r_{ij}^3} \left( s_i^x s_j^x + s_i^y s_j^y\right) 
         + \Delta \sum_{i>j} \frac{1}{r_{ij}^6}s_i^z s_j^z,%
\end{equation}
 We set $\Delta = 1.0$ and calculate $P^{Q}(S^{D}_2)$ for the horizontal cut in the parameter space $W = 0.1$  - Fig.~\ref{fig:disco_nonzero_delta}(a). We can see that a region with a positive dimerization parameter $\delta$ behaves very similarly to $\Delta = 0$ case, but the behavior for the negative $\delta$ is now different. Firstly, the fraction of states with $S^D_2 = 0$ is now more prominent (in the large $L$ limit it goes to $85 \%$). Secondly, the fraction of states with  $S^D_2 = 1, 2$ is now more modest. 
Moreover, the data for $\delta=-0.8$ plotted in Fig.~\ref{fig:disco_nonzero_delta}(b) indicates that this fraction vanishes with $L$ very quickly. For comparison, we plot a fraction of states that have $S^D_2 = 2$ and do not involve any bulk exchanges (having only $\ket{\pm}$ terms in the bulk) as a purple curve. It behaves very similarly to $P^Q(S^D_2=2) $ indicating a finite size effect, not extensive with the system size. 

The dependence of $P^Q$ on $\Delta \in [0, 1]$ may also be analyzed - compare Fig.~\ref{fig:disco_nonzero_delta}(c). It is apparent that a qualitative change occurs around $\Delta=0.1$. For larger $\Delta$ similar behaviour as for $\Delta=1$ is observed - with increasing system size the fraction of states with $S^D_2=0$ increases while those with long-range entanglement and $S^D_2=2$ goes down. For small $\Delta$ the trend for states with $S^D_2=0$ is reversed and the fraction of ``interesting'' states decays much slower.

\end{document}